# NON-INTEGER MULTIVERSE

Marvin Chester

Physics Department

University of California at Los Angeles

Los Angeles, California, USA

chester@physics.ucla.edu

**Abstract:**

In quantum mechanics physical processes procede by two different mechanisms. John von Neumann enumerated them as 1, the "discontinuous ... arbitrary changes by measurement," and 2, continuous evolution via the Schroedinger Equation. That the physical world does not obey a single overriding law - unitary evolution by the Schroedinger Equation - is philosophically disturbing to some. Others face it with equanimity. One narrative that preserves the findings of quantum mechanics yet does produce pure unitary evolution is that of the multiverse. Given below is the narrative by which Born's Rule emerges without pre-assigning to it the notion of probability. It requires that the number of universes in the multiverse not be enumerable!

## 1. The Stern-Gerlach Experiment Archetype

The archetypical experiment to study the matter of measurement is a Stern-Gerlach apparatus. Figure 1 shows a schematic of it. The pole faces of the magnet, one being edge-shaped and the other spread-shaped, provide, besides a magnetic field, $B_y$, a strong y-gradient in the strength of that field, $\partial B_y/\partial y$.

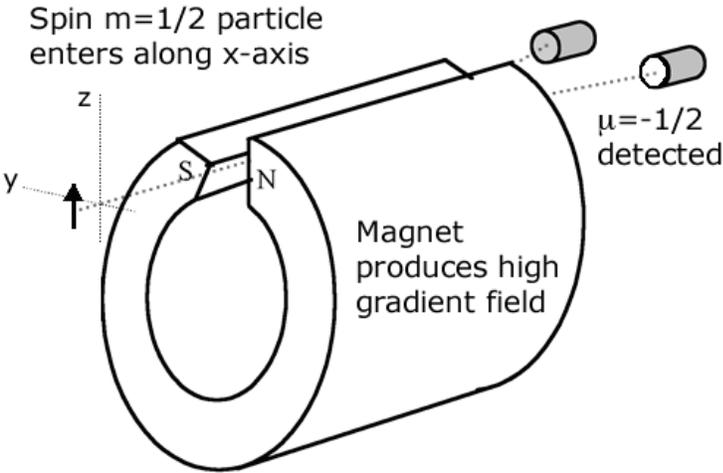

Figure 1.

Idealized Stern-Gerlach apparatus. The magnetic field y-gradient spatially separates the superposition states in the y-direction.

A spin is projected into a high gradient y-directed magnetic field region. No matter the orientation of the incoming particle's spin, its state is resolved into y-components by the apparatus. The strong field gradient acts differently on each of these y-spins physically pulling them apart. Thus the environment - the contoured magnet - produces a *measurement basis*. The basis is y-component spin. Call it µ. The incoming particle's state is separated physically into y-spin components, µ=±1/2.

The essentials of the Hamiltonian operator, $\hat{H}$, affecting the spin during its passage through the field, are given by

$$\hat{H} = \hat{p}^2 / 2M - C\hat{\mu}B_y - C\hat{\mu}\hat{y}\left(\partial B_y / \partial y\right) \qquad (1)$$

where $C\hat{\mu}$ is the magnetic moment operator, the particle's mass is M and its momentum operator is $\hat{p}$. The third term produces the coupling between position and spin that gives rise to the physical separation of the spin states. A wave packet centered at y=0, enters at time t=0. At any later time, t, a location measurable, which we call Y, has a value that characterizes the y-location of the ket. It changes with time as either plus $t^2$ ($Y_+$) or minus $t^2$ ($Y_-$) in concert with whether µ is plus or minus 1/2. Explicit details are given by Daniel E. Platt [1], by De Oliveira and Caldeira [2] and in the charming MIT experiment manual by J. Lieberman [3].

An m=1/2 particle enters the apparatus. It is in a state |m=1/2,Y=0>. The application of Schroedinger's equation to a spin passing through a region governed by the Hamiltonian of equation 1 yields

$$e^{-it\hat{H}/\hbar}\left|m=1/2,Y=0\right\rangle = \left|\mu=1/2,Y_+\right\rangle/\sqrt{2} + \left|\mu=-1/2,Y_-\right\rangle/\sqrt{2} \qquad (2)$$

Equation 2 is the state of the system approaching the detectors. But for each incoming particle no more than one of the two detectors registers. The figure shows the µ=-1/2 detector registering. In that run of the experiment, the state of the system found, is |µ=-1/2,Y->. But this state is not the one that evolved via the Schroedinger Equation. Put formally:

$$e^{-it\hat{H}/\hbar}|m=1/2, Y=0\rangle \neq |\mu=-1/2, Y_{-}\rangle \quad (3)$$

Mathematically this equation is clearly valid; it's deducible from equation 2. But it also represents an experimental finding: that the spin state revealed by the detector, |µ=-1/2,Y->, cannot have evolved unitarily from the initial spin state |m=1/2,Y=0>.

That the world evolves by two different processes was expressly noted by John von Neumann [4] long ago. He listed as number 1 the "arbitary changes by measurements" embodied here in equation 3. His number 2 referred to the unitary evolution via the Schroedinger Equation as expressed in equation 2.

That the physical world does not obey a single overriding law - unitary evolution by the Schroedinger Equation - is philosophically disturbing to some. Others confront it with equanimity. One narrative that preserves the mathematics of quantum mechanics yet does produce pure unitary evolution is that of the multiverse.

**2. The multiverse; a number of universes.**

There are a plethora of references pertaining to the multiverse idea. An exhaustive list on the subject is given by Sebens and Carroll [5]. Here is the application of this idea to the present experiment.

The system's measurement observable is µ. Everything outside of the spin - the system - is its environment. Collect together all of the commensurate measurement observables of the world outside of the system into the symbol, W. For notational convenience we also corral within W the spin-to-environment coupling that causes the physical separation in y of the two spins, µ=±1/2; W includes Y.

The essential insight of the multiverse viewpoint is that the observer - you - is part of the environment. Being also subject to the laws of nature you are exploring, you must include yourself in the computation. The observer is embedded in the W representing the environment of the experiment. This observer intends to repeat the same experiment many times.

In the first run of the experiment a spin enters with m=+1/2. The initial state of the system and its world is |m=1/2,W>. As before, the B-field gradient in the y-direction physically separates µ=+1/2 from µ=-1/2. Both measurement results µ=+1/2 <u>and</u> µ=-1/2 are produced. The superposition terms represent different worlds. Two 'you's materialize to replace the one existing before measurement, each attached to one of the measurement results. This is expressed as:

$$e^{-it\hat{H}/\hbar}|m=1/2,W\rangle = |\mu=1/2,W_+\rangle/\sqrt{2} + |\mu=-1/2,W_-\rangle/\sqrt{2} \qquad (4)$$

$W_+$ contains the 'you' that sees µ=+1/2 and $W_-$ contains the 'you' that sees µ=-1/2. The reason that you see µ=-1/2 is because 'you' are embedded in the world with that result.

Both the 'you' in $W_+$ and the one in $W_-$ continue on with the intent, implanted in them at W, to repeat the experiment. Doing so causes four 'you's to materialize; two in the worlds replacing $W_+$ and two in those replacing $W_-$. So by virtue of unitary evolution:

$$e^{-it\hat{H}/\hbar}|m=1/2,W_+\rangle = |\mu=1/2,W_{++}\rangle/\sqrt{2} + |\mu=-1/2,W_{+-}\rangle/\sqrt{2} \quad (5)$$

and

$$e^{-it\hat{H}/\hbar}|m=1/2,W_-\rangle = |\mu=1/2,W_{-+}\rangle/\sqrt{2} + |\mu=-1/2,W_{--}\rangle/\sqrt{2} \quad (6)$$

The environment, $W_{+-}$, contains the 'you' that saw the sequence µ=+1/2 followed by µ=-1/2. Doing the experiment 10 times will produce $2^{10}$ different 'you's each having observed a distinct sequence of ten pluses and minuses. Like + + + - - + - - + +. This is the history of observations made by one of the 'you's. It contains six pluses and four minuses. There are many histories that lead to six pluses and four minuses. The number of 'you's having seen p pluses and 10-p minuses is N(p) = 10!/p!(10-p)! which peaks sharply at p=5. The number of 'you's that have seen five pluses, p=5, is 252. The number that saw p=0 is only 1.

This distribution of histories arises for any one observer. So in an ensemble of different observers - say 1024 of them - we would expect 252 of them to report finding sequences with 5 pluses and 5 minuses. The distribution of findings in the real world is predicted to be N(p) = 10!/p!(10-p)! This becomes N(p) = N!/p!(N-p)! for N runs of the experiment.

This result is appealing. It matches experiment. The narrative offers a way to perceive probability as arising naturally from the unitary evolution of the wave function. We must accept that we are embedded together with our observations in a grander wave function. Any observation - measurement result - takes the observer with it. You are carried along with your measurement results. Thus there is no intrinsic 'you'. A myriad of 'you's has been growing interminably. By this device we have rescued the unitary evolution of the wave function of nature from von Neumann projection. All branches in a superposition survive. But being a part of the grand scheme of things, any observer is dragged along with his observations into one branch or another. Interminably.

There is a difficulty. Suppose we rotate the spin source so that the spin entering the apparatus is not in the z-direction but skew to it. Say at some angle, $\theta$, off the y-direction in the y-z plane. It enters in a state, $\eta=1/2$. By the above analysis this should not change the measurement results. But we know, in fact, that the measurement results are changed. For particles entering with a skew spin the results peaks at $p_{MAX} = N|\langle\mu=1/2|\eta=1/2\rangle|^2 = N\cos^2(\theta/2)$. That is the Born Rule result. Only for the special case above - initial spin in the z-direction, $\theta=\pi/2$ - is this quantity equal to $N/2 = |\langle\mu=1/2|m=1/2\rangle|^2 N$. So the narrative, put forth above, cannot be right. It gives wrong answers for any $\theta\neq\pi/2$.

The following narrative rectifies the shortcomings.

### 3. Non-integer number of universes.

An observer is embedded in environment W with system measurement observable, $\mu$ in a Stern-Gerlach apparatus. This observer - you - intends to repeat the same experiment many times. A spin enters with $\eta=+1/2$. The B-field gradient in the y-direction physically separates

µ=+1/2 from µ=-1/2. A detector interacts with the particle. You see a measurement result µ=+1/2 or µ=-1/2. Calling $|a|^2 := |\langle\mu=1/2|\eta=1/2\rangle|^2 = 1 - |b|^2$, the state of the system evolves to:

$$e^{-it\hat{H}/\hbar}|\eta=1/2,W\rangle = |\mu=1/2,W_+\rangle a + |\mu=-1/2,W_-\rangle b \tag{7}$$

As before several 'you's materialize. *We allow the amount of universes produced to be non-integer.* We postulate that there are $2|a|^2$ universes with ket |µ=+1/2,W_+>. For compactness call $2|a|^2:=f$. And there are $2|b|^2=2-f$ universes with ket |µ=-1/2,W_->.

Justification: Since the multiverse itself is quite incomprehensible why restrict it to integer numbers of universes? It's no less a prejudice than that there should be only one universe. [(incomprehensibility)$^2$= (incomprehensibility)].

Then there are f states |µ=+1/2,W_+> generated from |η=+1/2,W> in a single run of the experiment. Each environment, W_+ contains a 'you' that sees µ=+1/2. And there are $2|b|^2=2(1-|a|^2):=g$ universes with ket |µ=-1/2, W_-> where W_- contains the 'you' that sees µ=-1/2. Since f+g=2 there are a total number of 2 universes where there was 1 before. Both the 'you' in W_+ and the 'you' in W_- continue on with the intent, buried in them at W, to repeat the experiment. So, after another time t each of the states produces its own new universes:

$$e^{-it\hat{H}/\hbar}|\eta=1/2,W_+\rangle = |\mu=1/2,W_{++}\rangle a + |\mu=-1/2,W_{+-}\rangle b \tag{8}$$

and

$$e^{-it\hat{H}/\hbar}|\eta=1/2,W_-\rangle = |\mu=1/2,W_{-+}\rangle a + |\mu=-1/2,W_{--}\rangle b \tag{9}$$

From each of the f states, $|\eta=1/2,W_+\rangle$, issue f states, $|\mu=+1/2,W_{++}\rangle$, so there are now $(f)^2$ states, $|\mu=+1/2,W_{++}\rangle$. And from each of those same f states, $|\eta=1/2,W_+\rangle$ also issue g states, $|\mu=-1/2,W_{+-}\rangle$. So there are now (f×g) states, $|\mu=-1/2,W_{+-}\rangle$. Similar reasoning produces g×f states, $|\mu=1/2,W_{-+}\rangle$, and $g^2$ $|\mu=-1/2,W_{--}\rangle$ states from the g states, $|\eta=1/2,W_-\rangle$.

This analysis may be carried through for N runs of the experiment. Among the $2^N$ states spawned by the process one finds there are $N(p) = f^p g^{N-p} N!/p!(N-p)!$ of them that contain p pluses. Alternatively put, the number of 'you's that will find p pluses after N runs of the experiment is

$$N(p) = 2^N |a|^{2p} \left(1-|a|^2\right)^{N-p} N!/\left(p!(N-p)!\right) \qquad (10)$$

Thus the 'probability' of finding p, $N(p)/2^N$, peaks at $|a|^2$. The Born Rule says that the probability of finding the result $\mu=1/2$ given that the initial state was $\eta=1/2$ is $|a|^2 = |\langle\mu=1/2|\eta=1/2\rangle|^2$. So if we interpret $2|a|^2$ as the number of universes materializing with the property $\mu=1/2$ then we get what Born predicts; that in many experiments, $|a|^2$ is the probability of finding $\mu=1/2$.

So the narrative offered works. It gives correct answers. But in order to do so it invokes a non-denumerable amount of other universes. These spout forth interminably at every event that happens. I'm not sure that this is easier to accept conceptually than that two mechanisms govern quantum mechanics. Hopefully better narratives exist.

[1] Daniel E. Platt (1992) Am. J. Phys, 60, 306-308

[2] T.R. De Oliveira and A. O. Caldeira (2006) Phys. Rev. A 73, 042502

[3] J. Lieberman (1998) http://people.roma2.infn.it/~carboni/SternGerlach.pdf

[4] John Von Neumann (1955), "Mathematical Foundations of Quantum Mechanics", translated by Robert T. Beyer Princeton U Press p 351.

[5] Charles T. Sebens, Sean M. Carroll, (2015) "Self-Locating Uncertainty and the Origin of Probability in Everettian Quantum Mechanics" arXiv:1405.7577 [quant-ph]